\documentclass[useAMS,usenatbib,usegraphicx]{mn2e}

\title[Super-orbital Period in the High Mass X-ray Binary 2S 0114+650]
  {Super-orbital Period in the High Mass X-ray Binary 2S 0114+650
}
\author[S.A. Farrell et al.]
  {S.A.~Farrell$^1$,
  R.K. Sood$^1$
  and P.M. O'Neill$^2$ \\
  $^1$School of Physical, Environmental and Mathematical Sciences,
  UNSW@ADFA, Northcott Drive, Canberra ACT 2600, Australia \\
  $^2$Astrophysics Group, Imperial College London, Blackett Laboratory,
  Prince Consort Road, London SW7 2AZ, UK}
\date{Released 2005 Xxxxx XX}

\pagerange{\pageref{firstpage}--\pageref{lastpage}} \pubyear{2005}

\def\LaTeX{L\kern-.36em\raise.3ex\hbox{a}\kern-.15em
    T\kern-.1667em\lower.7ex\hbox{E}\kern-.125emX}

\begin{document}

\label{firstpage}

\maketitle

\begin{abstract}
We report the detection of a stable super-orbital period in the high-mass
X-ray binary 2S 0114+650.  Analyses of data from the \textit{Rossi
X-ray Timing Explorer (RXTE)} All-Sky Monitor (ASM) from 1996
January 5 to 2004 August 25 reveal a super-orbital period of 30.7
$\pm$  0.1 d, in addition to confirming the previously reported
neutron star spin period of 2.7 h and the binary orbital period of
11.6 d.  It is unclear if the super-orbital period can be ascribed
to the precession of a warped accretion disc in the system.
\end{abstract}


\section{Introduction}

The X-ray source 2S 0114+650 was discovered in 1977 during the
\textit{SAS 3} Galactic survey (Dower $\&$ Kelley 1977).  Its
optical counterpart, LSI $+65\degr$ 010 was identified as an 11 mag
B star with a broad H$\alpha$ emission line (Margon $\&$ Bradt
1977). Spectroscopic observations by Crampton, Hutchings $\&$ Cowley
(1985) from 1979 to 1985 led these authors to conclude that
optically this system resembled supergiant X-ray binaries similar to
Vela X-1, while its X-ray properties were more aligned to those of
Be X-ray binaries.  Reig et al. (1996) carried out further long term
optical and infrared studies in the period 1990-1995, and
reclassified the optical counterpart as a supergiant of spectral
type B1 and luminosity class Ia, at a distance of 7.0 $\pm$ 3.6 kpc.

The binary nature of 2S 0114+650, with a period of 11.59 d was also
optically confirmed by Crampton et al. (1985). Koenigsberger et al.
(1983) using \textit{Einstein, HEAO 1}, and \textit{OSO 8} data, and
Yamauchi et al. (1990) using \textit{Ginga} data had reported
compact object pulse periods of 894 s and 850 s respectively.
However, analysis of \textit{EXOSAT} data by van Kerkwijk $\&$
Waters (1989) and Apparao, Bisht $\&$ Singh (1991) failed to confirm
these periodicities. X-ray data from \textit{EXOSAT} and
\textit{ROSAT} spanning seven years from 1983 to 1990 were analysed
by Finley, Belloni $\&$ Cassinelli (1992), who discovered a 2.78 h
periodicity and attributed it to $\beta$ Cephei-type pulsations of
the donor star. Corbet, Finley $\&$ Peele (1999) analysed
\textit{Rossi X-ray Timing Explorer (RXTE)} All Sky Monitor (ASM)
data covering the period 1996 January 5 to 1998 July 3 and concluded
that the X-ray pulse properties at the period of 2.7 h could be best
explained if they emanated from accretion on to a slowly rotating,
highly magnetised neutron star. These authors also confirmed the
orbital period reported by Crampton et al. (1985), but with a
difference of 0.04 $\pm$ 0.01 d between the optical and X-ray
periods.  The 2.7 h spin period is by far the slowest known for an
X-ray pulsar, and compares with the other known periods between 69
ms and 1400 s.  Li $\&$ van den Heuvel (1999) have shown that such a
slow pulsar is possible within the lifetime indicated by the early
type companion if the pulsar was born as a magnetar with a magnetic
field of $\geq$ 10$^{14}$ G. The field would have decayed to a
present value of $\sim$(2-3) $\times$ $10^{12}$ G on a timescale of
$\leq {10^{5}}$ yr. In this paper we report the results of analyses
of \textit{RXTE} ASM data covering a period of more than 8 years,
which have revealed a strong modulation at 30.7 $\pm$ 0.1 d.

\section{Data and Analysis}
Archived data from the ASM in the public domain\footnote
{http://heasarc.gsfc.nasa.gov/docs/xte/xtegof.html} (Levine et al.
1996) obtained for the period MJD 50087 (1996 January 5) to MJD
53242 (2004 August 25) were used in these analyses.  The ASM
consists of three wide-angle Scanning Shadow Cameras (SSC's) each
with a $6\degr$ $\times$ $90\degr$ field of view and spatial
resolution 3' $\times$ 15', equipped with proportional counters
sensitive to X-rays in the nominal energy range 1.5 -- 12.0 keV (A.
M. Levine 2004, private communication).  The spectral data are
binned into three energy bands corresponding to 1.5 -- 3.0 (soft,
A), 3.0 -- 5.0 (medium, B), and 5.0 -- 12.0 keV (hard, C).
\textit{RXTE} is in a low-earth circular orbit of inclination
$23\degr$, and period $\sim$95 min.  The ASM performs 90 s pointed
observations (dwells) of individual sources.  $\sim$80$\%$ of the
sky is covered every orbit, with a duty cycle of $\sim$40$\%$.

Data from the ASM are available in two forms:  count rates from
individual 90 s dwells, and one-day averages for a source.  Though
the daily averages are more suitable for probing long-term flux
variability, as the errors on flux measurements are reduced compared
to the individual dwell measurements (Corbet 2003), we used the 90 s
dwell data in order to investigate the spin period at a higher time
resolution.  The SSC3 data were not used because of that detector's
performance degradation over time. Linear trends were subtracted
from the SSC1 and SSC2 light curves before and after a gain
adjustment that took place on MJD 51548 (Corbet 2003).  Data points
which had count rate errors of 1$\sigma$ greater than the mean error
were removed from the light curve. The correlation between flux and
error in the ASM data points is very weak.  The error values are
determined from a model that represents the coded aperture analysis
of the instrument, where it is presumed that the sources and the
response functions (i.e. shadow patterns) are known (A. M. Levine
2005, private communication). A plot of the flux against errors
showed no apparent correlation between the two, confirming that our
screening method is not biased against times when the source flux is
low. In addition, power spectra using a filtered light curve were
very similar to those obtained using a weighted light curve. The
reduced light curve for 2S 0114+650 for the 8.5 years of ASM data
for the SSC1 and SSC2 detectors is shown in Fig. 1.  The mean
intensity is 0.3 counts s$^{-1}$ (4 mCrab) and the mean error is
0.99 counts s$^{-1}$.

\begin{figure}
\begin{center}
\includegraphics[width=8cm]{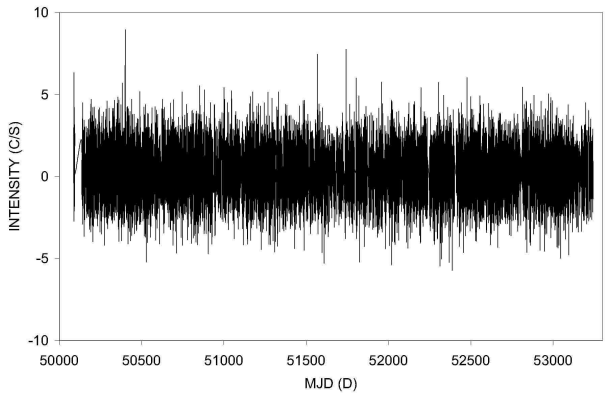}
\caption{The individual dwell light curve for 2S 0114+650 obtained
with the ASM.  Data points with count rate errors of 1$\sigma$
greater than the mean error have been removed. (1 c/s = 13.2 mCrab
in the energy range 1.5 - 12 keV).}
\end{center}
\end{figure}

After carrying out background subtraction and barycentric
corrections we performed a Lomb-Scargle periodogram (LSP) analysis
(Lomb 1976; Scargle 1982) for the SSC1 and SSC2 data.  As the False
Alarm Probability of a given peak in a LSP power spectrum depends on
the sampling pattern of the data (Horne $\&$ Baliunas 1986), we used
Monte Carlo simulations to determine the 99.9$\%$ white noise
significance levels (e.g. Kong, Charles $\&$ Kuulkers 1998). The
resulting power spectrum for the complete energy range 1.5 -- 12.0
keV is shown in Fig. 2.  The strongest peak in the spectrum is at
11.60 $\pm$ 0.02 d FWHM.  This value is in agreement with the value
of the orbital period as deduced by Crampton et al. (1985) from
optical observations.  As noted above, a previous analysis of the
first two years of ASM data by Corbet et al. (1999) had revealed a
value of 11.630 d as the optimum value for the folded light curve.
We find that the orbital period has remained stable at 11.60 $\pm$
0.02 d during the 8.5 yr observation period discussed in this paper.
Fig. 2 also shows a marginally significant peak at 2.73 h, which has
previously been associated with the pulse period of the neutron
star.  In their analysis, Corbet et al. (1999) had noted the broad
profile of this peak. They had also seen a subsidiary peak at 25.65
h which they had provisionally attributed to daily variations in
background levels. This peak is not significant in our data (Fig.
2).  We have made a detailed investigation of the occurrence of
peaks around 24 h in ASM data for several celestial sources, and
ascribe them primarily to spectral leakage of low frequency power
present in the light curves (Farrell, O'Neill $\&$ Sood 2005).
\begin{figure}
\begin{center}
\includegraphics[width=8cm]{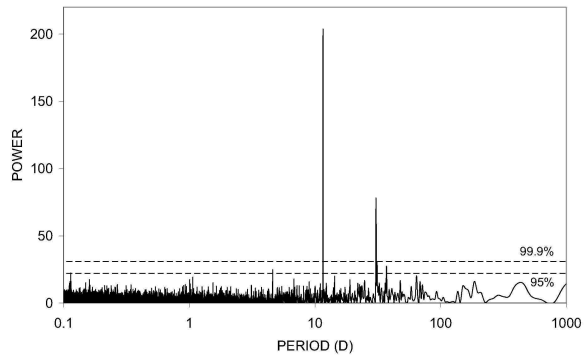}
\caption{Lomb-Scargle power spectrum of the individual dwell
barycentre-corrected, background subtracted light curve for 2S 0114+650
for 8.5 yr of ASM data.  The 99.9$\%$ and 95$\%$
significance levels are indicated by the dashed lines.}
\end{center}
\end{figure}

We have examined the evolution of the neutron star spin period in 2S
0114+650 using ASM data over the period MJD 50087 -- 53242. The
$\sim$8.5 yr light curve was split into 14 smaller overlapping data
sets each 400 d in length. These sections were then analysed
separately using the LSP technique. The results are shown in Fig. 3.
While the $\sim$2.7 h period was not significantly detected in all
the data sub-sets, we confirm the neutron star spin values
previously derived by Corbet et al. (1999), Hall et al. (2000), and
Bonning $\&$ Falanga (2005) for overlapping time intervals.  A
decrease in the spin period of the secondary was first noted by Hall
et al.  Our detailed analysis shows two episodes of torque reversal,
each on a time scale of $\sim$400 days.  These episodes are not
accompanied by a significant change in the X-ray flux from the
source.

\begin{figure}
\begin{center}
\includegraphics[width=8cm]{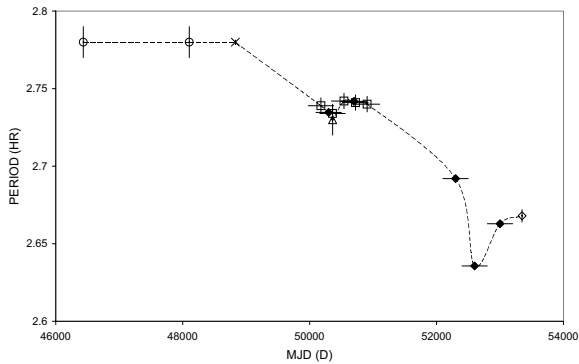}
\caption{Evolution of the neutron star spin period in 2S 0114+650.
The data points are as follows:  open circle - Finley et al. 1992,
open square - Corbet et al. 1999, open triangle - Hall et al. 2000,
cross - Finley et al. 1994, open diamond - Bonning $\&$ Falanga
2005, filled diamond  - present work.}
\end{center}
\end{figure}

In addition to the previously reported periodicities in the ASM data
as described above, Fig. 2 shows a strong peak at $\sim$30 d period.
This peak is well above the 99.9$\%$ white noise significance level
as determined from Monte Carlo simulations.  We also simulated a
light curve, using the same sampling as our data, with modulations
at both the 2.7 h pulse and the 11.6 d orbital periods.  A
periodogram calculated from this light curve showed no power at
around 30 d, confirming that the observed 30.7 d peak is not a
result of spectral leakage from the pulse or orbital periods.  A
follow-up $\chi$$^{2}$ epoch-folding search of the data yielded a
best-fit period of 30.7 $\pm$ 0.1 d FWHM.  The analysis gave an
ephemeris of MJD 50108.2 ($\pm$ 0.4) + 30.7 ($\pm$ 0.1)N, where N is
the cycle number.  The phase variability of 2S 0114+650 over this
period is shown in Fig. 4.  Phase zero is defined as the modulation
minimum, and the modulation has full amplitude of 66 $\pm$ 25$\%$.
We interpret this modulation as a previously unreported
super-orbital period in the 2S 0114+650 system (Farrell, Sood $\&$
O'Neill 2004)\footnote{Following the announcement of this discovery
in ATEL $\#$283, it was brought to our attention by Dr. J. Wilms
that this periodicity had been seen for an unspecified length of ASM
data set by Benlloch (PhD thesis, Univ. of T\"{u}bingen, 2004),
albeit at a much lower significance level.}.

\begin{figure}
\begin{center}
\includegraphics[width=8cm]{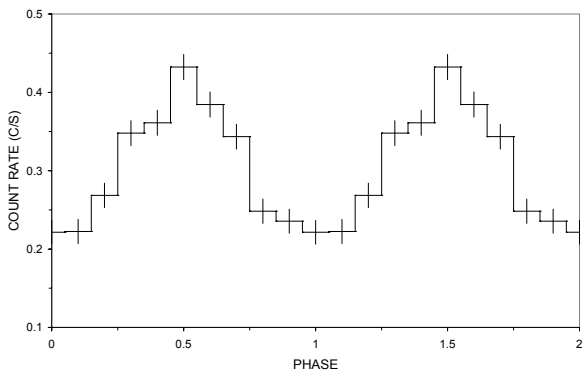}
\caption{Light curve for ASM data for 2S 0114+650, folded over the
30.7 d super-orbital period at epoch T(0) = MJD 50108.2, with two
cycles shown for clarity.}
\end{center}
\end{figure}

In order to confirm that there was no contamination in the 2S
0114+650 data from other neighbouring known sources, we performed a
similar analysis on ASM data from the nearby 3.6 s X-ray pulsar 4U
0115+634, and the LMXB pulsar 4U 0142+614. While the LSP showed a
strong peak at $\sim$24 h (discussed above) in 4U 0115+634, no
significant power was found in either source at any of the periods
associated with 2S 0114+650.

The 2S 0114+650 data for the three energy bands were analysed
separately to investigate the energy dependence of the temporal
variation of the X-ray flux. The power spectra for the individual
energy bands are shown in Fig. 5. There are no significant
periodicities in the 1.5 -- 3.0 keV (A) band. The 11.6 d orbital
period is highly significant in the 3.0 -- 5.0 keV (B) band, with
the 30.7 d super-orbital period appearing at the $>$ 99$\%$
significance level. Both the orbital period, and the super-orbital
period stand out strongly in the 5.0 -- 12.0 keV (C) band, while the
spin period marginally exceeds the 99$\%$ white noise significance
line. The A, B and C channel light curves folded at the 30.7 d
super-orbital period are shown in Fig. 6.

We also checked to see if the C$/$B and B$/$A hardness ratios
revealed any periodic modulation.  We generated power spectra after
discarding $\sim$15$\%$ of the hardness ratio data points which had
B and A values close to zero.  The power spectra showed no
significant modulation at any period. Note that Corbet et al. (1999)
also found that the ASM data are not of a sufficient quality to
establish whether the hardness ratio varies with the orbital period.

It is useful also to investigate explicitly whether the X-ray
spectrum varies as a function of the phase of the super-orbital
period because these flux variations might be owing to changes in
absorption. Using the A, B and C channel light curves, folded at
30.7 d (see Fig. 6), we calculated the folded time-series of the
C$/$(A + B) hardness ratio. First, we fitted the time-series with a
constant hardness, which yielded a satisfactory fit
($\chi$$^2$/D.O.F.  = 16.3/9;  null hypothesis probability 0.061).
We then added a sine curve to the model with the same phase as the
observed 30.7 d flux modulation.  The 90$\%$ confidence interval
($\Delta$$\chi$$^2 = 2.71$) of the amplitude was found to be from
$-$0.4 to 0.4. This confidence interval corresponds to a 90$\%$
upper limit on the $\emph{peak-to-peak}$ amplitude of 0.8, either in
phase or $\pi$ radians out-of-phase with the intensity modulations.
We conclude that the ASM data are not of sufficient quality to
confirm or preclude variable absorption as the mechanism behind the
super-orbital modulation.

\begin{figure}
\begin{center}
\includegraphics[width=8.1cm]{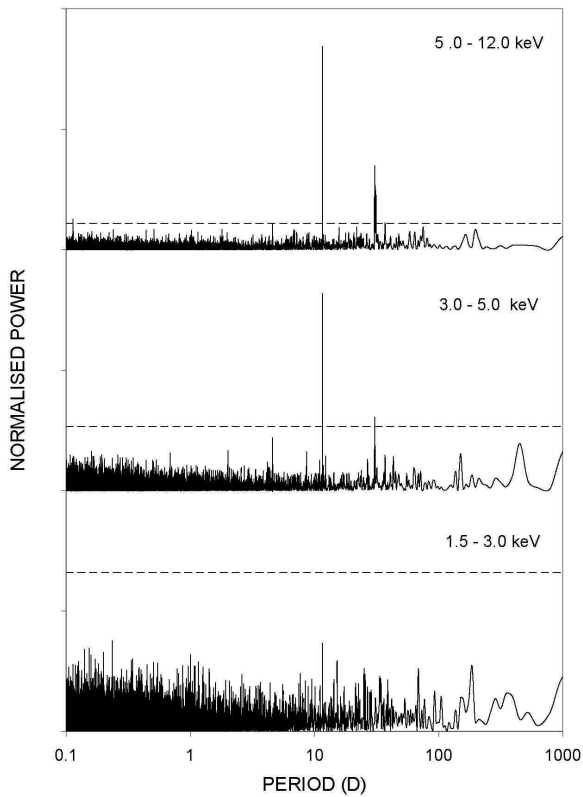}
\caption{Lomb-Scargle power spectra for 8.5 yr data for 2S 0114+650,
for the three energy bands of the ASM.  The 99$\%$ significance
levels are indicated by the dashed lines.}
\end{center}
\end{figure}

\begin{figure}
\begin{center}
\includegraphics[width=8cm]{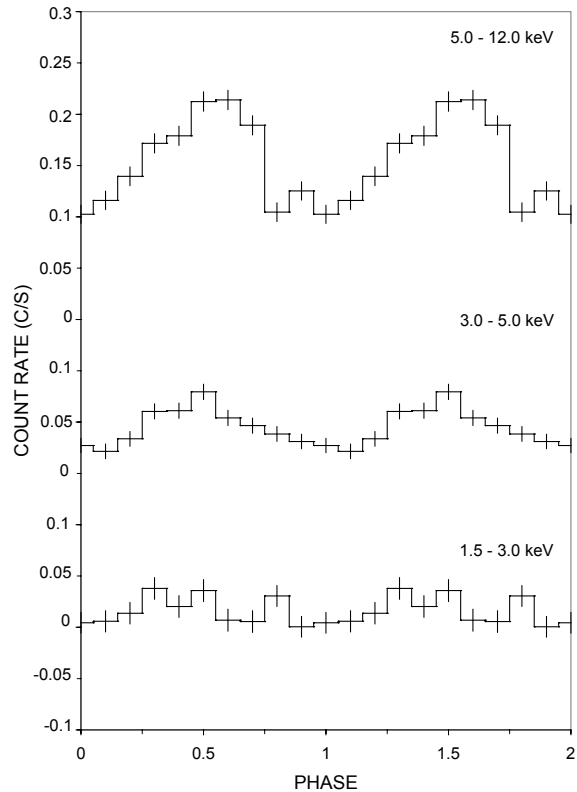}
\caption{Light curves for the A (1.5 - 3.0 keV), B (3.0 - 5.0 keV),
and C (5.0 - 12.0 keV) energy bands folded over the 30.7 d
super-orbital period at epoch T(0) = MJD 50108.2.}
\end{center}
\end{figure}

We also confirmed that the 11.6 d orbital period modulation was not
distorting the results obtained for the super-orbital modulation.
The orbital period was removed by subtracting a sine curve fitted to
the light curve folded at 11.6 d. This procedure produced only a
slight amplitude shift of the super-orbital modulation from 0.20
c$/$s (66$\%$) p-p to 0.19 c$/$s (64$\%$) p-p.

In order to determine whether the 30.7 d super-orbital period is
persistent and coherent, we split the light curve into four
$\sim$800 d sub-sets covering the ranges MJD 50134 -- 50934, MJD
50935 -- 51733, MJD 51735 -- 52534, and MJD 52535 -- 53242. Each
sub-set was analysed using the LSP technique, and then folded at the
30.7 d period with an epoch of MJD 50108.2. Fig. 7 shows the folded
light curve sections. The super-orbital period is not significantly
present in the power spectrum of the first two years of the
$\it{RXTE}$ ASM data (the MJD 50134 -- 50934 data set), although
there does appear to be a very small amplitude modulation in the
folded light curve. This would explain why the 30.7 d period did not
show up in the power spectrum of these data in the work of Corbet et
al. (1999). The modulation does however show up clearly in the other
three data sets with minimal variation in phase, although the
amplitudes vary considerably.

We have examined the evolution of both the super-orbital and orbital
periods using the same method described earlier for the spin period,
with the results shown in Fig. 8. The spin evolution trend has been
overlaid for comparison. The super-orbital period appears to have
varied slightly throughout the mission lifetime, although the
relatively large error bars prohibit a definitive conclusion. It is
quite clear however that the 30.7 d period has not been evolving in
concert with the spin period. As expected the orbital period has not
been varying over time, although the location of the peaks in the
LSP power spectra appear to shift slightly in unison with a similar
periodic shift in the super-orbital period.

\begin{figure}
\begin{center}
\includegraphics[width=8cm]{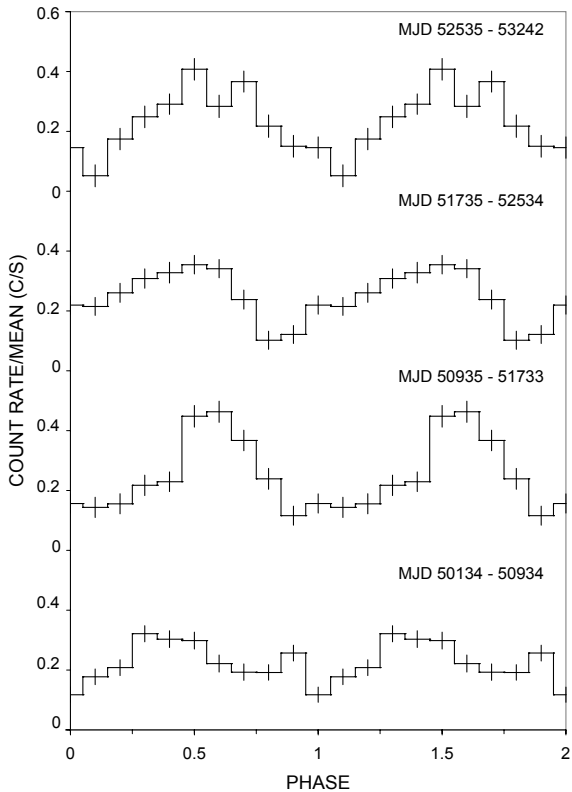}
\caption{800 d light curve sections folded over the 30.7 d
super-orbital period at epoch T(0) = MJD 50108.2.}
\end{center}
\end{figure}

\begin{figure}
\begin{center}
\includegraphics[width=8cm]{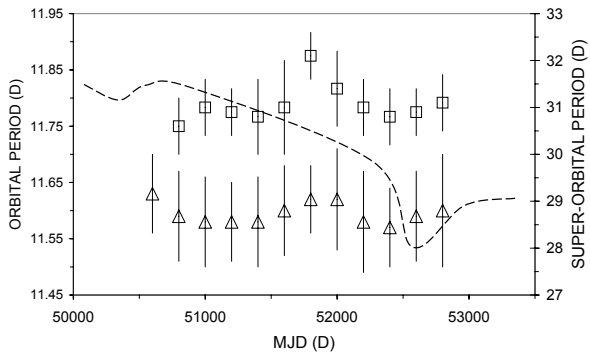}
\caption{Evolution of the orbital (open triangle) and super-orbital
(open square) periods, with the neutron star spin-up trend from Fig
3 overlaid for comparison (dashed line).}
\end{center}
\end{figure}

\section{Discussion}

Super-orbital (long) periods are known to exist in more than 20
high-mass and low-mass X-ray binaries, and their values range from
24 to 600 d (e.g. Ogilvie $\&$ Dubus 2001). Her X-1, LMC X-4, and SS
433 have super-orbital periods which are very stable (Paul $\&$
Kitamoto 2002), and we have demonstrated above that 2S 0114+650 is
the fourth such system with a stable period.  The mechanism behind
these periods is generally not well-understood, though the
precession of a radiation-warped accretion disc modulating the X-ray
flux from the compact object is the favoured model (e.g. Clarkson et
al. 2003). Ogilvie $\&$ Dubus (2001) have shown that radiation
driven warping gives a coherent picture of super-orbital periods,
but that the model does not explain the phenomenon in all X-ray
binaries.  Other models for super-orbital periods include (i)
variations in the rate of Roche-lobe overflow caused by stellar
pulsations of the donor (Weiler et al. 1992), (ii) the precession of
the magnetic axis of the compact object (Tr\"{u}mper et al. 1986),
(iii) a triple star system (Chou $\&$ Grindlay 2001), and (iv)
variations in the location of an accreting hot-spot (Rutten, van
Paradijs $\&$ Tinbergen 1992). We now discuss the likely phenomenon
leading to a super-orbital period in 2S 0114+650.

2S 0114+650 shares luminosity (3.4 $\times$ $10^{35}$ erg s$^{-1}$
at 3 -- 20 keV) and spectral characteristics with systems that
contain pulsars in \textit{wind-driven} HMXRBs, though some
uncertainty remains over its optical spectral characteristics.  The
hydrogen column density is estimated at (1.3 -- 9.4) $\times$
$10^{22}$ cm$^{-2}$, with the lower value being an upper limit to
interstellar absorption (Reig et al. 1996). It is therefore clear
that significant and variable absorption takes place in the vicinity
of the X-ray source. However, the structure of the absorbing region
is not certain.  It may be in the form of a precessing accretion
disc, or a dense circumstellar environment.  There are arguments
against the existence of an accretion disc in 2S 0114+650.  The
formation of an accretion disc is expected only when the Keplerian
radius r$_{k}$ of orbiting matter of specific angular momentum
$\it{l}$ is larger than the magnetospheric radius r$_{m}$ of the
neutron star of mass M$_{x}$, where $r_{k}$ =
$l^{2}$/$\it{GM}$$_{x}$, (Nagase 1989). The value of the stellar
wind velocity (1200 km s$^{-1}$) far exceeds the calculated upper
limit value ($\sim$280 km s$^{-1}$) necessary for the formation of
an accretion disc in 2S 0114+650. Additionally, the HeII 4686{\AA}
line emission, a common feature of the inner regions of a heated
disc in X-ray binaries, has not been observed from this system
(Koenigsberger et al. 2003).

The fractional amplitude of the variations of the X-ray flux arising
from the precession of an accretion disc should be largest in the
low energy bands of the ASM, because the soft X-rays are absorbed
more than the hard X-rays.  As shown in Fig. 5, this modulation is
only present in the higher energy bands. Variations in the rate of
mass loss from the donor could in principle explain the 30.7 d
periodicity.  Such a variation should manifest itself in a
correlated variation at optical wavelengths. Hudec (1978) studied
the mean B band brightness of the system over a 50 yr interval. The
brightness was constant at 12.3 $\pm$ 0.2 mag from 1928 to 1977. The
radial velocity measurements of Crampton et al. (1985) in the range
3700 -- 5100 {\AA}, taken over more than six years, show no evidence
of modulation at the super-orbital period. Reig et al. (1996) used
the HeI 6678 {\AA} absorption feature for radial velocity
determination of 2S 0114+650, concluding that optical variability on
the time scale of months to years was not a feature of 2S 0114+650.
Variations in mass loss rate are also not expected to show precision
timing at the super-orbital period when considered over a period of
several years.

More significant is the fact that there is one other report in the
literature of a periodicity in 2S 0114+650 at $\sim$30 d. This
relates to the work of Beskrovnaya (1988) who carried out UBVRI
polarimetry of the optical counterpart LSI $+65\degr$ 010 over a two
month period in 1986.  He confirmed the variability of the
normalised Stokes parameters Q$/$I and U$/$I over the 11.6 d orbital
period, and stated that ``there also seems to be an auxiliary source
of variability, with a characteristic time scale of roughly a
month''. The donor star has a strongly ionised wind, and this could
be a dominant source of the linear polarisation, resulting from
single electron scatter. Note that the presence of an accretion disc
cannot be ruled out on the basis of the polarisation observations,
as a correlation between the polarisation parameters and the orbital
phase is expected in X-ray binaries that contain accretion discs
(e.g. Phillips $\&$ Meszaros 1986). Alternatively, the precession of
the magnetic axis of the neutron star could conceivably result in
varying irradiation of the circumstellar material, leading to a
natural explanation for the 30-d variation in the polarisation, and
the observed super-orbital period. This mechanism was initially
proposed for Her X-1, in which the 1.24 s pulse profile changes with
the system's super-orbital phase (Tr\"{u}mper et al. 1986).  These
authors showed that the necessary external torque supplied by an
accretion disc in Her X-1 to cause forced precession  was deficient
by a factor of 10$^{6}$.  Similar consideration must apply to 2S
0114+650, even though a correlation between the pulse and
super-orbital periods is difficult to establish because of the
limitations of the ASM data for this source, and due to the weakness
of its flux density (Fig. 1 and 2). When internal magnetic stresses
are large enough, the free precession of a rotating neutron star
with an oblique magnetic field has been shown to be inevitable
(Wasserman 2003). However, the predicted precession period for the
neutron star in  2S 0114+650, with a slow spin period of 2.7 h,
turns out to be orders of magnitude larger than 30.7 d for accepted
values of the parameters involved.

The picture is further complicated if one plots the spin$/$orbital
behaviour of 2S 0114+650 on the Corbet diagram (Corbet 1986), which
shows the relation between the spin period $P_{\rmn{s}}$ and the
orbital period $P_{\rmn{o}}$ for accreting neutron stars. The
expected relationship (Corbet 1986) is explained by the neutron star
being in a state of quasi-equilibrium in which the Alfven and
co-rotation radii of the neutron star are approximately equal.  The
size of the Alfven radius depends on the density of the
circumstellar material, which in turn will depend on the binary
separation.  The observed relationship is shown in Fig. 9 for
various classes of X-ray binaries (Bildsten et al. 1997). Trends are
evident for Be transients, and for Roche-lobe filling supergiants.
There is no correlation for under-filled Roche-lobe supergiants. In
this scheme, 2S 0114+650 belongs to the latter class of binaries,
wherein an accretion disc is not present (Fig. 9).  The system may
be spinning up to its equilibrium state. If on the other hand, the
system was near the equilibrium state, then the neutron star would
need to possess a present value of the magnetic field of $>$
$10^{\rmn{13}}$ G (Waters $\&$ van Kerkwijk 1989).
\begin{figure}
\begin{center}
\includegraphics[width=8cm]{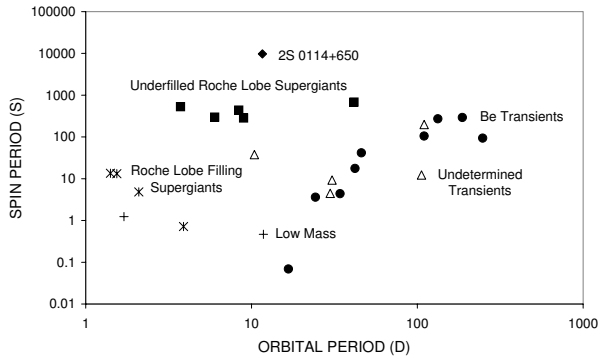}
\caption{The spin$/$orbital periods of the accreting neutron stars
(the Corbet diagram) with 2S 0114+650 included, with the other data
points taken from Bildsten et al. (1997).}
\end{center}
\end{figure}

Chou $\&$ Grindlay (2001) have shown that a long period ($\sim$176
d) in 4U 1820-30 is produced by a hierarchical triple outer
companion star modulating the eccentricity of an inner binary orbit.
4U 1820-30 consists of a white dwarf $/$ neutron star inner system
with $P_{\rmn{orb,inner}}$ $\approx$ 685 s, with a third star
orbiting the inner system with a period of $\sim$1.1 d.  If this
model (Mazeh $\&$ Shaham 1979) is applied to 2S 0114+650 taking
$P_{\rmn{orb,inner}}$ = 11.6 d, it yields a long period of $\gg$
30.7 d.

\section{Conclusions}

We have identified a third stable period in the high mass X-ray
binary 2S 0114+650 of value 30.7 $\pm$ 0.1 d, making this system the
fourth known X-ray binary to possess a stable super-orbital period. Such a
period is usually classified as originating from modulation of the
X-ray flux by a warped precessing accretion disc. However, the
observational evidence for the presence of an accretion disc in this
system is not conclusive. Optical polarisation and spectroscopy
studies, and radio observations are in progress to better understand
the properties of this complex system.

\section{Acknowledgements}

This research has made use of data obtained through the High Energy0
Astrophysics Science Archive Research Centre Online Service,
provided by the NASA/Goddard Space Flight Centre.  We thank Alan
Levine of MIT for discussions on the ASM instrument, and Stefan
Dieters of the University of Tasmania for his advice on all matters
related to the \textit{RXTE} mission.

\label{lastpage}

\end{document}